\documentclass[american,aps,pra,reprint,superscriptaddress,showpacs]{revtex4-1}
\usepackage[T1]{fontenc}
\usepackage[latin9]{inputenc}
\usepackage{color}
\usepackage{babel}
\usepackage{amsthm}
\usepackage{amsmath, amssymb}
\usepackage{bm}
\usepackage{bbm}
\usepackage{pxfonts,txfonts}
\usepackage{tgpagella}
\usepackage[T1]{fontenc}

\usepackage{amsmath}
\usepackage{amssymb}
\usepackage{graphicx}
\usepackage[unicode=true,pdfusetitle,
 bookmarks=true,bookmarksnumbered=false,bookmarksopen=false,
 breaklinks=false,pdfborder={0 0 0},backref=false,colorlinks=true]
 {hyperref}
\hypersetup{
 linkcolor=urlcolor,citecolor=urlcolor,urlcolor=urlcolor}

\makeatletter
\@ifundefined{textcolor}{}
{%
 \definecolor{BLACK}{gray}{0}
 \definecolor{WHITE}{gray}{1}
 \definecolor{RED}{rgb}{1,0,0}
 \definecolor{GREEN}{rgb}{0,1,0}
 \definecolor{BLUE}{rgb}{0,0,1}
 \definecolor{CYAN}{cmyk}{1,0,0,0}
 \definecolor{MAGENTA}{cmyk}{0,1,0,0}
 \definecolor{YELLOW}{cmyk}{0,0,1,0}
 }
\theoremstyle{plain}

\theoremstyle{plain}

\ifx\proof\undefined
\newenvironment{proof}[1][\protect\proofname]{\par
\normalfont\topsep6\p@\@plus6\p@\relax
\trivlist
\itemindent\parindent
\item[\hskip\labelsep
\scshape
#1]\ignorespaces
}{%
\endtrivlist\@endpefalse
}
\providecommand{\proofname}{Proof}
\fi
\theoremstyle{plain}

\theoremstyle{plain}

\usepackage{txfonts}
\usepackage{braket}
\definecolor{urlcolor}{rgb}{0,0,0.7}

\makeatother

\providecommand{\lemmaname}{Lemma}
\providecommand{\propositionname}{Proposition}
\providecommand{\theoremname}{Theorem}
\providecommand{\corrolaryname}{Corrolary}

\begin{document}

\title{Quantum optics and frontiers of physics: The third quantum revolution}

\author{Alessio Celi}

\affiliation{ICFO -- Institut de Ci\`encies Fot\`oniques, The
Barcelona Institute of Science and Technology, Av. C.F. Gauss 3,
08860 Castelldefels, Spain}

\author{Anna Sanpera}

\affiliation{ICREA -- Instituci\'o Catalana de Recerca i Estudis
Avan\c cats, Lluis Companys 23, 08010 Barcelona, Spain}

\affiliation{Departament de F\'isica, Universitat Aut\`onoma
Barcelona, 08193 Bellaterra, Spain}

\author{Veronica Ahufinger}

\affiliation{Departament de F\'isica, Universitat Aut\`onoma
Barcelona, 08193 Bellaterra, Spain}

\author{Maciej Lewenstein}

\email{maciej.lewenstein@icfo.es}

\affiliation{ICFO -- Institut de Ci\`encies Fot\`oniques, The
Barcelona Institute of Science and Technology, Av. C.F. Gauss 3,
08860 Castelldefels, Spain}

\affiliation{ICREA -- Instituci\'o Catalana de Recerca i Estudis
Avan\c cats, Lluis Companys 23, 08010 Barcelona, Spain}

\begin{abstract}

The year 2015 was the International Year of Light. It marked,
however, also the 20th anniversary of the first observation of
Bose-Einstein condensation in atomic vapors by Eric Cornell, Carl
Wieman and Wolfgang Ketterle. This discovery can be considered as
one of the greatest achievements of quantum optics that has
triggered an avalanche of further seminal discoveries and
achievements. For this reason we devote this essay for the focus
issue on ``Quantum Optics in the International Year of Light'' to
the recent revolutionary developments in quantum optics at the
frontiers of all physics: atomic physics, molecular physics,
condensed matter physics, high energy physics and quantum
information science. We follow here the lines of the introduction
to our book ``Ultracold atoms in optical lattices: Simulating
quantum many-body systems'' \cite{Lewen2012}. The book, however,
was published in 2012, and many things has happened since then --
the present essay is therefore upgraded to include the latest
developments.
\end{abstract}


\maketitle

The achievement of \index{Bose-Einstein condensation (BEC) ! alkalines}
Bose-Einstein condensation (BEC) in dilute gases in
1995~\cite{Anderson95,Davis95,Bradley95} marks the beginning
of a new era. For the so-called AMO community ---comprising
atomic, molecular, optics and quantum optics--- it became soon
evident that condensed matter physics, i.e., degenerate quantum
many body systems could be at reach. The condensed matter
community remained at that stage more skeptical. They argued that,
at the very end, what was achieved experimentally was a regime of
weakly interacting Bose gases; the domain thoroughly investigated
by the condensed matter theorists in the 50's and 60's
\cite{Mahan93,Fetter71}. For solid state/condensed matter
experts the fact that the AMO experiments dealt with confined
systems of finite size and typically inhomogeneous densities was a
technical issue rather than a question of fundamental importance.
Nonetheless, the Nobel foundation awarded its yearly prize in 2001
to E. A. Cornell, C. E. Wieman and W. Ketterle ``for the
achievement of Bose-Einstein condensation in dilute gases of
alkali atoms, and for early fundamental studies of the properties
of the condensates'' \cite{Cornell02,Ketterle02}. Today, from
the perspective of some years, we see that due to the efforts of
the whole community these fundamental studies have enriched
amazingly the standard ``condensed matter'' understanding  of static
and dynamical properties of weakly interacting Bose gases
\cite{Pitaevskii03}.

From the very beginning the AMO community continued their efforts
to push the BEC physics towards new regimes and new challenges.
The progress in these directions was indeed spectacular, and at
the beginning of the third millennium became clear to both, AMO
and condensed matter communities, that indeed we are entering a
truly new quantum era with unprecedented possibilities of control
of many body systems. In particular, it became clear that the
regime of strongly correlated systems may be reached with
ultracold atoms and/or molecules. Few years after the first
observation of BEC, in 1999 \index{Fermi systems  degenerated gas}
atomic degenerate Fermi gases were achieved
~\cite{DeMarco99,Truscott01,Schreck01}  paving the way toward the
observations of \index{Fermi systems  superfluidity} Fermi
superfluidity.  This is described, in the weak interaction limit
by the \index{Bardeen-Cooper-Schrieffer (BCS)  theory}
Bardeen--Cooper--Schrieffer theory (BCS) \cite{Fetter71}, and by the
so called \index{BCS-BEC ! crossover} BEC-BSC crossover in the
limit of strong correlations. For recent reviews reporting the
large activity in this field see \cite{Giorgini08} and
\cite{Bloch08}. Even earlier, following the seminal proposal by
Jaksch {\it et al.} \cite{Jaksch98}, Greiner {\it et al.}
\cite{Greiner02} observed the signatures of the quantum phase
transition from the \index{SF-MI phase transition} superfluid to
the, so called Mott insulator state for bosons confined in an
optical lattice.

Nowadays, ultracold atomic and molecular systems are at the
frontiers of modern quantum physics, and are considered to provide
the most controllable systems to study many body physics. It is
believed that these systems will also find highly nontrivial
applications in quantum information, and quantum metrology and
will serve as powerful quantum simulators. At the theory level,
the broadness of the different fields touched by ultracold atoms
has lead to a ``grand unification''  where AMO, condensed matter,
nuclear physics, and even high energy physics theorists are
joining efforts ---for reviews see \cite{Lewenstein07} and
\cite{Bloch08}---.

After the first quantum revolution at the beginning of the XX
century, the second one associated with the name of John Bell and
the experimental quest for non-locality of quantum mechanics
together with the experimental control over single, or few
particle systems ---see in particular Alain Aspect's introduction
to \cite{Bell04}---, we are witnessing the third one: the quantum
revolution associated to the control over macroscopic quantum
systems and the rise of quantum technologies.
Indeed this quantum revolution is not limited to the ultracold atoms but 
includes other platforms like NV-centers \cite{review-NVc} and 
superconducting qubits \cite{review-scq} where quantum optics also
plays a central role.

\section{Cold atoms from a historical quantum optical perspective}
\label{sec_1_2}

The third quantum revolution does not arrive completely
unexpected: the developments of atomic physics and quantum optics
in the last 30 years have inevitably led to it. In the seventies
and eighties of the XX century, atomic physics was a very well
established and respectful area of physics. It was, however, by no
means a ``hot'' area. On the theory side, even though one had to
deal with complex problems of many electron systems, most of the
methods and techniques were already developed. The major problems
that were considered concerned mainly questions related on
optimisation of these methods. These questions were reflecting an
evolutionary progress, rather than a revolutionary search for
totally new phenomena. On the experimental side, quantum optics at
that time was entering its Golden Age. The development of laser
physics and nonlinear optics led in 1981 to the Nobel prize for
A.L. Schawlow and N. Bloembergen ``for their contribution to the
development of laser spectroscopy''. On the other hand, studies of
quantum systems at the single particle level culminated in 1989
with the Nobel prize for H.G. Dehmelt and W. Paul ``for the
development of the ion trap technique'', shared with N.F. Ramsey
''for the invention of the separated oscillatory fields method and
its use in the hydrogen maser and other atomic clocks''.

Theoretical quantum optics was born in the 60'ties with the works
on quantum coherence theory developed the 2005 Noble prize winner,
R.J. Glauber \cite{Glauber63a,Glauber63b}, and with the
development of the laser theory in the late 60'ties  by H. Haken
and by M.O. Scully and W.E. Lamb (Nobel laureate of 1955). In the
70'ties and 80'ties, however, theoretical quantum optics was not
considered to be a separate, established area of theoretical
physics. One of the reasons of this state of the art, was that
indeed quantum optics at that time was primarily dealing with
single particle problems. Most of the many body problems of
quantum optics, such as laser theory, or more generally optical
instabilities \cite{Walls06}, could have been solved either
using linear models, or employing  relatively simple versions of
the mean field approach. Perhaps the most sophisticated
theoretical contributions concerned understanding of quantum
fluctuations and quantum noise \cite{Walls06,Gardiner04}.

This situation has drastically changed due to the unprecedented
level of {\it quantum engineering}, i.e. preparation,
manipulation, control and detection of quantum systems developed
and achieved by atomic physics and quantum optics. There are
several seminal discoveries that have triggered these changes:
\begin{itemize}

\item The cooling and trapping of atoms, ions and molecules have reached regimes of very low
temperatures (today down to nano-Kelvin!) and precision that were
considered unattainable just two decades ago. These developments
were recognized by the Nobel Foundation in 1997, who awarded the
Prize to S. Chu \cite{Chu98}, C. Cohen-Tannoudji
\cite{Cohen-Tannoudji98} and W.D. Phillips \cite{Phillips98} ``for
the development of methods to cool and trap atoms with laser
light''. \index{cooling  laser} Laser cooling and mechanical
manipulations of particles with light \cite{Metcalf01} was
essential for development of completely new areas of atomic
physics and quantum optics, such as atom optics \cite{Meystre01},
and for reaching new territories of precision metrology and
quantum engineering.

\item Laser cooling combined with \index{cooling evaporative} evaporation
cooling allowed, in 1995, the experimental observation
Bose-Einstein condensation (BEC) \index{Bose-Einstein condensation (BEC) ! alkalines}
\cite{Anderson95,Davis95,Bradley95}, a phenomenon predicted by S. Bose and A. Einstein
more than 70 years earlier. As we said before, these experiments marked evidently the birth of
a new era. E.A. Cornell and C.E. Wieman \cite{Cornell02} and W. Ketterle \cite{Ketterle02} received
the Nobel Prize in 2001, ``for the achievement of Bose-Einstein condensation in dilute gases of
alkali atoms, and for early fundamental
 studies of the properties of the condensates''. This  was truly a breakthrough moment, in which
 ``atomic physics and quantum optics
has met condensed matter physics'' \cite{Ketterle02}.  Condensed
matter community at that time remained, however, still reserved.
After all, BEC was observed in weakly interacting dilute gases,
where it is very well described by the mean field
\index{Bogoliubov-de Gennes (BdG)  approach} Bogoliubov-de Gennes
theory \cite{Pitaevskii03}, developed for homogeneous systems in
the 50'ties.

\item The study of \index{quantum correlations} quantum correlations and entanglement. The seminal
theoretical works of the late A. Peres \cite{Peres93,Peres96}, the
proposals of quantum cryptography by C. H. Bennett and G. Brassard
\cite{Bennett84}, and A. K. Ekert \cite{Ekert91}, the quantum
communication proposals by C. H. Bennett and S. J. Wiesner
\cite{Bennett92}  and C. H. Bennett, G. Brassard, C. Cr{\' e}peau,
R. Jozsa, A. Peres, and W. K. Wootters \cite{Bennett93}, the
discovery of the quantum factorizing algorithm by P. Shor
\cite{Shor94}, and the \index{quantum computation  ions proposal}
quantum computer proposal by J. I. Cirac and P. Zoller
\cite{Cirac95} gave birth to the field of  quantum information
\cite{Bouwmeester00,Zoller05a} (for the latest version of the
quantum information road map see \cite{qurope}). These studies,
together with the rapid development of the theory have led to an
enormous progress in our understanding of quantum correlations and
entanglement and, in particular, how to prepare and use entangled
states as a resource. The impulses from quantum information enter
nowadays constantly into the physics of cold atoms, molecules and
ions, and stimulate new approaches. It is very probable that the
first quantum computers will be, as suggested already by Feynman
\cite{Feynman82,Feynman86}, computers of special purpose --
\index{quantum simulator ! basic concepts} quantum simulators (QS)
\cite{Cirac04}, that will efficiently simulate quantum many body
systems that otherwise cannot be simulated using ``classical''
computers \cite{Zoller05a}.

\item Optical lattices and \index{Feshbach resonances} Feshbach resonances. The physics of ultracold atoms entered the area of strongly correlated systems
with the  seminal proposal of Jaksch {\it et al.} \cite{Jaksch98}
on how to achieve the \index{SF-MI phase transition} transition
superfluid-Mott insulator transition in cold atoms using optical
lattices. The proposal was based on the bosonic Mott insulator
transition proposed in condensed matter~\cite{Fisher89}, but the
authors of this paper were in fact motivated by the possibility of
realizing quantum computing with cold atoms in a lattice.
Transition to the Mott insulator state was supposed to be an
efficient way of preparing a quantum register with a fixed
number of atoms per lattice site. Entanglement between atoms could
be achieved via controlled collisions~\cite{Jaksch99}. The
experimental observation of the superfluid-Mott insulator
transition by the Bloch--H{\''a}nsch group \cite{Greiner02} was
without any doubts a benchmark  in the studies of strongly
correlated systems with ultracold atoms \cite{Bloch04}. Several
other groups have observed bosonic superfluid-Mott insulator
transitions in pure Bose systems \cite{Stoferle04a}, in
\index{disordered systems ! bosons} disordered Bose systems
\cite{Fallani07}, \index{Bose-Fermi mixtures ! model} in
Bose-Fermi \cite{OspelkausS06,Gunter06}, and \index{Bose-Bose
mixtures} Bose-Bose mixtures \cite{Thalhammer08}. The possibility
to change the collision properties of ultracold atoms by means of
tuning the Feschbach resonances of the atomic species has been
another tool of inestimable value. Such a tool has lead to the
\index{Mott insulator (MI)  fermions} fermionic Mott insulator
\cite{Jordens08,Schneider08}. Also, \index{Mott insulator (MI) !
molecules} Mott insulator state of molecules have been created
\cite{Volz06}, as well as  bound repulsive pairs of atoms (i.e.\
pairs of atoms at a site that cannot release
their repulsive energy due to the band structure of the spectrum in  the lattice) have been observed \cite{Winkler06}.\\

\item Cold \index{trap ! ions} trapped ions were initially investigated to
realize the Cirac-Zoller quantum gate \cite{Cirac95}, and to
attempt to build an scalable \index{quantum computation ions
proposal} quantum computer using a bottom-up
approach~\cite{SchmidtKaler03,Home09}. 
Recent progress in this research line can be found in \cite{blatt2012quantum}.
A new stream of ideas using
cold ions for quantum simulators have recently appeared. First
proposals have shown that ion-ion interactions mediated by phonons
can be manipulated to implement various spin models
\cite{Mintert01,Wunderlich02,Porras04a}, where the spin states
correspond to internal states of the ion. In tight linear traps,
one could realize the spin chains with interactions decaying as
(distance)$^{-3}$, i.e. as in the case of dipole-dipole
interactions. Such spin chains may serve as \index{quantum neural
network} quantum neural network models and may be used for
\index{quantum information ! adiabatic processing} adiabatic
quantum information processing \cite{Pons07,Braungardt07}. More
interestingly, ions can be employed as \index{quantum simulator !
ion systems} quantum simulators, both in 1D and in 2D, where the
ions form self-assembled \index{Coulomb ! crystal} Coulomb
crystals \cite{Porras06a}. First steps towards the experimental
realization of these ideas has been 
reported~\cite{Friedenauer08}. 
These results have been recently extended in \cite{Kim09,Kim10,Britton11,islam2013emergence,Jurcevic14,Richerme14}
Alternatively to spins, one could
look at \index{phonon ! ions} phonons in ion self-assembled
crystals; these are also predicted to exhibit interesting
collective behaviour from Bose condensation to strongly correlated
states \cite{Porras04b,Deng08}. Trapped ions may be used also to
simulate \index{spin-boson model} mesoscopic spin-boson models
\cite{Porras08}. Combined with optical traps or ion microtrap
array methods can be used to simulate a whole variety of spin
models with tunable interactions in a wide range of spatial
dimensions and geometries \cite{Schmied08}. Finally, they are
promising candidates for the realization of many-body interactions
\cite{Bermudez09}. All these theoretical proposals and the
spectacular progress in the experimental trapped ion community
pushes trapped ions physics to be soon used as widely as cold
atoms to mimic condensed matter physics and beyond, as for instance in \cite{martinez2016real}.
\end{itemize}

\section{Quantum simulators}
\label{sec_1_2.5}

Before we proceed we would like to make a short {\it detour} to
explain in a little more detail the concept of {\it quantum
simulators}. XX century was the age of information and computers.
But, even supercomputers have their limitations -- as we well
know for instance their ability to predict weather, a
paradigmatic classical chaotic phenomenon, is very restricted. For
this reason already in classical computer science the concept of
special purpose computers was developed. These ``classical
simulators'' are not like universal classical computers -- they can
only simulate or calculate certain restricted class of models
describing Nature. The best simulations of classical disordered
systems, such as spin glasses, are nowadays obtained with such
computers of special purpose -- ``classical simulators''.

The situation is even more dramatic in quantum physics and
chemistry. We know very well that an universal quantum computer
will revolutionize our technology in the future. Unfortunately,
this future does not seem to be very close at hand, and definitely
concerns tens of years.  However, quantum computers with a special
purpose, i.e. quantum simulators, exist since already 5 years (cf.
\cite{RunnersUp10}).

Various quantum phenomena such as high-$T_c$ superconductivity or
quark confinement are still awaiting universally accepted
explanations because of the computational complexity of solving
the simplified theoretical models designed to capture the relevant
physics. Richard Feynman suggested in 1982
\cite{Feynman82,Feynman86}, solving such models by
''quantum simulation''. He pointed out that it might be possible
sometimes to find a simpler and experimentally more accessible
system to mimic the quantum system of interest. Feynman's idea was
motivated by the complexity of classical simulations of quantum
systems. Suppose we want to study a system of $N$ spins 1/2; then
the dimension of the Hilbert space is $2^N$, and the number of
coefficients we need to describe the wave function of the system
may be in principle just as large. As $N$ grows, this number
quickly becomes larger than the number of atoms in the universe,
so classical simulations are evidently impossible. Of course, in
practice this number may be very much reduced by using clever
representations of the wave functions, but in general there are
many quantum systems, which are very hard to simulate classically.
The modern concept of quantum simulators is not exactly the same
as that of Feynman. Since condensed matter systems are typically
very complex, theorists construct simplified models. In
this simplification, symmetries of the original problem are kept
intact, and the concept of universality is used: different
Hamiltonians with similar symmetry properties belong to the same
universality classes, i.e. they exhibit the same phase transitions
and have the same critical exponents. Unfortunately, even these
simplified models are often difficult to understand. A
paradigmatic example of such a situation concerns high-$T_c$
superconductivity of cuprates, where it is believed that the basic
physics is captured by an array of weakly coupled 2D Hubbard
models for electrons, i.e. spin-1/2 fermions. Even this simple
model reduced to one 2D plane does not allow for accurate
treatment, and there is much controversy concerning, for instance,
the phase diagram or the character of the transitions. The role of
quantum simulators, as proposed in many quantum information
projects, will be to simulate simple models, such as 2D Hubbard
models, and obtain a better understanding of them, rather than try
to simulate the full complexity of the condensed matter.

A ``working'' definition of a quantum simulator could be as follows
\cite{Buluta09,Lewen2012}:
\begin{itemize}
\item A quantum simulator is an experimental system that mimics
a simple model, or a family of simple models, of condensed
matter (or high-energy physics, or quantum chemistry...).
\item  The simulated models have to be of some relevance for
applications and/or our understanding of the challenges of
the above-mentioned areas of physics, 
and of the new challenges in Physics, Chemistry and Biology that
are still to be identified.
\item The simulated models should be computationally
intractable for classical computers. Note that this statement
may have two meanings: i) an efficient (scalable to large
system size) algorithm to simulate the model might not
exist, or might not be known; ii) the efficient scalable
algorithm may be known, but the size of the simulated
model is too large to be simulated under reasonable time
and memory restrictions. The latter situation, in fact, occurs
with classical simulations of the Bose or Fermi Hubbard
models as compared to their experimental quantum
simulators. There might also be exceptions to the general
rule. For instance, it is desirable to realize quantum
simulators to simulate and to observe novel, hitherto only
theoretically predicted phenomena, even though it might
be possible to simulate these phenomena efficiently with
present-day computers. Simulating and observing is more
than just simulating.
\item A quantum simulator should allow for broad control of the
parameters of the simulated model, and for control of the
preparation, manipulation and detection of the states of
the system. In particular, it is important to be able to set the
parameters in such a way that the model becomes tractable
using classical simulations. This provides the possibility of
validating the quantum simulator.
\end{itemize}

Fueled by the prospect of solving a broad range of long-standing
problems in strongly-correlated systems, the tools to design,
build, and implement QSs \cite{Feynman82,Buluta09,Lewen2012} have
rapidly developed and are now reaching very sophisticated
levels~\cite{RunnersUp10}. There exist many proposals and
already many realizations of quantum simulators employing
ultracold atoms and molecules in traps and in optical lattices
(cf.
\cite{Anderlini07,Y.-J.Lin09a,Y.-J.Lin09b,Bakr10,Trotzky10,Jordens10,Liao10,Simon11,Struck11,VanHoucke11,Bloch-last,Spielman-last},
 for a review
see also \cite{Bloch08}), probing quantum dynamics (cf.
\cite{trotzky-probing-2011,cheneau-light-cone-like-2012}),
employing ultracold trapped ions (cf.
\cite{Friedenauer08,Kim10,Islam11,Lanyon11,Barreiro11,Gerritsma10,Gerritsma11},
for recent reviews see \cite{Johanning09,Schneider12}), atoms in
single or in arrays of cavities (cf. \cite{Baumann10}), ultracold
atoms/ions in arrays of traps, ultracold atoms near
nano-structures and multidimensional plasmonic traps \cite{thompson2013coupling,gonzalez2015subwavelength,stehle2011plasmonically,julia2013engineering}, photons
\cite{lanyon_experimental07,lanyon_experimental08,belgiorno_hawking10,kitagawa_observation11,ma_quantum11},
arrays of quantum dots, circuit quantum electrodynamics (QED) and
polaritons (cf.
\cite{Greentree06,Hartmann06,Angelakis07,Wang09a,Koch10}, for a
recent review see \cite{Angelakis16}), artificial lattices in
solid state \cite{Polini},   nuclear magnetic resonance (NMR)
systems \cite{Somaroo99,Tseng99,Du10,Alvarez10} and
superconducting qubits.

In addition to our book \cite{Lewen2012} there exist excellent
recent reviews, in particular in the focus issue of Nature Physics
Insight
\cite{Trabesinger2012,nphys-zoller,nphys-bloch,nphys-blatt,nphys-aspuru,nphys-koch}.
 At the current pace, it
is expected that we will soon reach the ability to finely control
many-body systems whose description is outside the reach of a
classical computer. For example, modeling interesting physics
associated with a quantum system involving 50-100 spin-1/2
particles -- whose general description requires $2^{50} - 2^{100}
\approx 10^{15} - 10^{30}$ amplitudes -- is out of the reach of
current classical supercomputers, but perhaps within the grasp of
a quantum simulators.

Obviously, the real-world implementations of a quantum simulation
will always face experimental imperfections, such as noise due to
finite precision instruments and interactions with the
environment.  The quantum simulator -- as envisioned by Feynman --
is fundamentally an {\it analog} device, in the sense that all
operations are carried out continuously. However, errors in an
analog device (also continuous, like temperature in the initial
state, or the signal-to-noise ratio of measurement) can propagate
and multiply uncontrollably~\cite{Kendon10}.  Indeed, Landauer, a
father of the studies of the physics of information, questioned
whether quantum coherence was truly a powerful resource for
computation because it required a continuum of possible
superposition states that were ``analog'' in
nature~\cite{Landauer96}.  In contrast to digital quantum
simulators, where the {\it error correction schemes} might be in
principle applied, the question of validation, calibration and
reliability of analog quantum simulators is very subtle and very
''hot'' \cite{HaukeROPP}.

\section{Cold atoms and quantum optics at the frontiers of  physics}
\label{sec_1_3}

Quantum optics and  physics of cold atoms touches nowadays, among others, common frontiers of contemporary physics
with condensed matter physics, quantum many body physics,  nuclear physics, quantum field theory, high energy physics,
and even astrophysics. In particular, many important challenges of the latter  disciplines can be addressed within
cold atoms and quantum optical wizardry:

\begin{itemize}

\item {\it 1D systems.} The role of \index{fluctuations ! quantum} quantum fluctuations and \index{correlation ! 1D systems} correlations is particularly important in 1D. The theory of 1D systems is very well developed due to the existence of exact methods such as \index{Bethe ansatz} Bethe Ansatz and \index{quantum inverse scattering theory} quantum inverse scattering theory  (cf.\ \cite{Essler05}),
powerful approximate approaches, such as \index{bosonization}
bosonization, or \index{conformal field theory (CFT)} conformal field
theory ---cf.\ \cite{Giamarchi04}---, and efficient
computational methods, such as \index{density matrix
renormalization group (DMRG)} density matrix Renormalization
group technique (DMRG) ---cf.\ \cite{Schollwoeck05}---.
There are, however, many open experimental challenges that have
not been so far directly and clearly realized in condensed matter,
and can be addressed with cold atoms ---for a review see
\cite{Cazalilla04, Cazalilla11}---. Examples include atomic Fermi, or
Bose analogues of spin-charge separation, or more generally
observations of microscopic properties of \index{Luttinger ! liquid}
Luttinger liquids \cite{Recati03,Paredes03,Kollath05}.
Experimental observations of the 1D gas in the  deep
\index{Tonks-Girardeau gas} Tonks-Girardeau regime by Paredes et
al. \cite{Paredes04} ---see also \cite{Moritz03},
\cite{Kinoshita04}, \cite{Stoferle04a} and
\cite{Laburthe04}--- were the first steps in this
direction. The recent achievement of Tonks regime in 2008
employing \index{dissipative ! processes} dissipative processes (two
body losses) is perhaps the first experimental example of how to
control a system by making use of its coupling to the environment
\cite{Syassen09,Durr08,Durr09,Garcia-Ripoll09}. See also the
recent progress towards deep Tonks-Girardeau regime in
\cite{Haller10}.

\item{\it Spin-boson model}. \index{spin-boson model} A two-level system coupled to a bosonic reservoir is a
 paradigmatic model both, in quantum dissipation theory in quantum optics and as quantum phase transition in condensed matter where it is termed as 
 the spin-boson model ---for a review see \cite{Leggett87}---. It has also been proposed \cite{Recati05} 
 that an atomic quantum dot, i.e., a single atom in a tight optical trap
coupled to a superfluid reservoir via laser transitions, may
realize this model. In particular, atomic quantum dots embedded in
a 1D Luttinger liquid \index{Luttinger ! liquid} of cold bosonic
atoms accomplishes a spin-boson model with Ohmic coupling, which
exhibits a dissipative phase transition and allows to directly
measure atomic \index{Luttinger ! parameter} Luttinger parameters.
Also, it has been shown that in the low-energy limit the system of a driven quantum spin coupled to a bath of ultracold fermions
can be mapped onto the spin-boson model with an Ohmic bath \cite{Knap13}

\item{\it 2D systems}. \index{low dimensional systems ! 2D ! continuous symmetry} According to the celebrated \index{Mermin-Wagner-Hohenberg theorem}
Mermin-Wagner-Hohenberg theorem, 2D systems with continuous
symmetry do not exhibit long range order at temperatures $T>0$. 2D
systems may, however, undergo \index{Kosterlitz-Thouless-Berezinskii (KTB) ! transition}
Kosterlitz-Thouless-Berezinskii transition (KTB) to a state in
which the correlations decay is algebraic, rather than
exponential. Although KTB transition has been observed in liquid
Helium \cite{Bishop78,Bishop80}, its microscopic nature
(binding of vortex pairs) has never been seen. Recent experiments
~\cite{Hadzibabic06,Hadzibabic08,Clade09,Hadzibabic10,Tung10,Hung11,Chomaz15} make
an important step in this direction.


\item{\it Hubbard and spin models}. \index{Hubbard model ! basic concepts} \index{spin ! models} 
Many very important examples of strongly correlated states in condensed matter physics are realized in various types of Hubbard models
\cite{Auerbach94,Essler05}. While Hubbard models in condensed matter are ``reasonable caricatures'' of real systems, 
ultracold atomic gases in optical lattices allow to achieve practically perfect realizations of a whole variety of Hubbard models \cite{Jaksch05}. 
Similarly, in certain limits Hubbard models reduce to various spin models; again cold atoms and ions allow for practically perfect realizations of such spin models ---see for instance
\cite{Dorner03,Duan03,Santos04,Garcia-Ripoll04} and \cite{Kim09,Simon11,Britton11,Jurcevic14,Richerme14}---. 
Specifically, spin-spin interactions between neighboring atoms can be implemented by bringing the atoms together on a 
single site and carrying out \index{collisions ! control} controlled collisions \cite{Sorensen99,Jaksch99,Mandel03}, 
\index{interactions ! exchange} on-site exchange interactions \cite{Anderlini07} or \index{interactions ! superexchange} 
superexchange interactions \cite{Trotzky08}. Moreover, one can use such realizations as quantum simulators to mimic specific condensed matter models.

\item{\it Disordered systems: Interplay localization-interactions}. \index{disordered systems ! interplay localization-interactions} Disorder plays a central
role in condensed matter physics, and its presence leads to
various novel types of effects and phenomena. One of the most
prominent quantum signatures of disorder is \index{Anderson !
localization} Anderson localization \cite{Anderson58} of the
wave function of single particles in a random potential. In cold
gases, controlled disorder, or pseudo--disorder might be created
in atomic traps, or optical lattices by adding an optical
potential created by \index{potential ! speckle} speckle radiation,
or by superposing several \index{optical lattices ! superlattice} lattices with
incommensurate periods of spatial oscillations
\cite{Damski03,Roth03b}. Other proposed methods are the
admixture of different atomic species randomly trapped in sites
distributed across the sample and acting as impurities ~\cite{Gavish05}, or the use of an
inhomogeneous magnetic field which modifies randomly, close to a
\index{Feshbach resonances} Feshbach resonance, the scattering
length of the atoms~\cite{Gimperlein05}. In fact, recently, 
experimental realization of Anderson localization of
matter waves has been reported in a non-interacting BEC of
$^{39}$K in a pseudo-random potential \cite{Roati08}, and in
a small condensate of $^{87}$Rb in a truly random potential in the
course of expansion in a one-dimensional waveguide
\cite{Billy08}. Three dimensional Anderson localization has also been reported
for a spin-polarized atomic Fermi gas of $^{40}$K \cite{Kondov11} and for $^{87}$Rb 
ultracold atoms \cite{Jendrzejewski12}. The experiment reported in \cite{Billy08}, as well as the
appearance of the \index{effective mobility edge} effective
mobility edge due to the finite correlation length of the speckle
induced disorder, was precisely predicted in
\cite{Sanchez-Palencia07}. The interplay between disorder
and interactions is a very active research area, also in ultracold
gases. For attractive interactions, disorder might destroy the
possibility of superfluid transition (''dirty'' superconductors)
\index{dirty ! superconductors}. Weak repulsive interactions play a
delocalizing role, whereas very strong ones lead to Mott type
localization \cite{Mott68a}, and insulating behavior. In the
intermediate situations there exist a possibility of
\index{quantum phases ! delocalized metallic} delocalized ``metallic'' phases.
Cold atoms in optical lattices should allow to study the crossover
between Anderson-like (Anderson glass) \index{Anderson ! glass} to Mott type \index{Bose systems !
Bose-glass}(Bose glass) localization. First experimental signatures of
a Bose glass \cite{Fallani07,Pasienski10,DErrico14}, of the
Anderson glass crossover~\cite{Deissler10} and of glassy behaviour in binary atomic 
mixtures \cite{Gadway11} have been reported. Theoretical
predictions \cite{Schulte06,Kuhn05,Bilas06,Paul05} indicate
signatures of Anderson localisation in the presence of weak
nonlinear interactions and quasi-disorder in  BEC. One expects in
such systems the appearance of a novel \index{Lifshits ! glass}
Lifshits glass phase \cite{Lugan07}, where bosons condense in
a finite number of states from the low energy tail \index{Lifshits ! tail} 
of the single particle spectrum. Very recently, the emergence of 
a disorder-induced insulating state \cite{Kondov15} and the observation of many-body 
localization \cite{Schreiber15} in interacting fermions have been reported. 

\item{\it Disordered systems: spin glasses}. \index{disordered systems ! spin glass}\index{spin ! glass} Since the seminal papers of Edwards
and Anderson \cite{Edwards75}, and Sherrington and
Kirkpatrick \cite{Sherrington75} the question about the
nature of the spin glass ordering has attracted a lot of attention
\cite{Mezard87,Fisher86,Bray87}. The two competing pictures:
the \index{replica symmetry breaking picture} replica symmetry
breaking picture of G. Parisi, and the droplet model
\index{droplet model} of D.S. Fisher and D.A. Huse are probably
applicable in some situations, and not applicable in others.
Ultracold atoms in optical lattices might contribute to resolve
this controversy by, for instance, studying independent copies
with the same disorder, the so--called replicas. A measurement
scheme for the determination of the disorder-induced correlation
function between the atoms of two independent replicas with the
same disorder has been proposed \cite{Morrison08}. Cold
atomic physics might also add understanding of some quantum
aspects, like for instance behavior of \index{Ising model ! spin glass}
Ising spin glasses in transverse fields ---i.e. in a truly quantum
mechanical situation \cite{Sanpera04,Ahufinger05}. More recently, the  realization of the  Dicke model with ultracold atoms in a 
 single mode cavity \cite{Baumann10}, stimulated recently intensive  studies of 
the connection between multi-mode Dicke models with random couplings and the spin glass physics ~\cite{Strack11,Gopalakrishnan11}. 
Also,  the possibility of addressing NP versions of the number partitioning were mentioned and investigated in  this context~\cite{Rotondo15,Raventos15}.

\item{\it Spin glasses and D-Wave computers}. In fact spin models are paradigms of multidisciplinary science. 
They are most relevant 
for various fields of physics, reaching from condensed matter to high energy 
physics, but they also find several applications beyond the physical sciences. 
In neuroscience, brain functions are modeled by interacting spin systems, 
going back to the famous Hopfield  model of associative memory~\cite{Hopfield82}. 
This directly relates  to computer and information sciences, where pattern 
recognition or error-free coding can be achieved using spin models~\cite{nishimori}. 
Importantly, many optimization problems, like number partitioning or the famous 
traveling salesman problem, belonging to the class of NP-hard problems, can be 
mapped onto the problem of finding the ground state of a specific spin 
model~\cite{Barahona82,Lucas14}. This implies that solving a spin model itself is a 
task for which no general efficient classical algorithm is known to exist.
A controversial development, 
supposed to provide also an exact numerical understanding of spin glasses, 
regards the D-Wave machine. These devices employing arrays of superconducting junctions, were recently introduced on the market. 
What they do is in  fact that they solve (i.e. find the ground state of)  classical spin glass models by using the, so called {\it quantum annealing approach}. 
Within this approach, if the system is trapped in a local minimum of energy, it can leave this configuration via quantum tunneling.   
Many researchers agree that D-Wave computers are genuine quantum simulators,
because despite decoherence and coupling to a thermal bath, they are consistent with performing quantum annealing \cite{smolin2013classical,wang2013comment},
and with open quantum system dynamics \cite{boixo2013experimental},
 but the underlying mechanisms are not clear, 
and it remains an open question whether these machines provides a speed-up advantage over the best classical algorithms~\cite{Troyer14,Katzgraber14,Denchev15}. 
All this triggers interest in alternative quantum systems, in particular quantum optical systems, designed to solve general spin models via quantum simulation. 
A noteworthy system for this goal are trapped ions: Nowadays, spin systems of trapped ions are available in many 
laboratories~\cite{Friedenauer08,Kim09,Kim10,Britton11,Jurcevic14,Richerme14}, and adiabatic state preparation, similar to quantum annealing, is experimental state-of-art.

\item{\it Disordered systems: Large effects by small disorder}. There are 
many examples of such situations. In classical statistical
physics a paradigm is the \index{Ising model ! in random fields}
random field Ising model in 2D ---that looses spontaneous
magnetization at arbitrarily small disorder---. In quantum physics
the paradigmatic example is \index{Anderson ! localization} Anderson
localization, which occurs at arbitrarily small disorder in 1D,
and should occur also at arbitrarily small disorder in 2D. Cold
atomic physics may address these questions, and, in fact, much
more --- cf. \cite{Wehr06}, \cite{Niederberger08a}
and \cite{Niederberger09}, where disorder breaks the
continuous symmetry in a spin system, and thus allows for long
range ordering---.

\item{\it High $T_c$ superconductivity}. \index{high $T_c$ superconductivity} 
Despite many years of research, opinions
on the nature of high $T_c$ superconductivity still vary quite
appreciably \cite{Claeson03}. It is, however, quite
established ---cf.\ contribution of P.W. Anderson in
\cite{Claeson03}--- that understanding of the
2D Hubbard model in the, so called,
\index{t-J model} $t-J$ limit
\cite{Spalek77,Chao77,Auerbach94,Spalek07} for two component
(spin 1/2) fermions provides at least part of the explanation. The
simulation of these models are very hard and numerical results are
also full of contradictions. Cold fermionic atoms with spin (or
pseudospin) $1/2$ in optical lattices might provide a
\index{quantum simulator ! fermionic atoms} quantum simulator to resolve these
problems \cite{Hofstetter02} ---see also
\cite{Koetsier06}---. First  experiments with both
''spinless''\index{Fermi systems ! polarized gas}, i.e.\ polarized, as well as
\index{Fermi systems ! unpolarized gas} spin $1/2$ unpolarized ultracold
fermions \cite{Stoferle04a,Stoferle06} have been realized;
particularly spectacular are the recent observations of a
\index{Mott insulator (MI) ! fermions} fermionic Mott insulator
state \cite{Jordens08,Schneider08}. It is also worth noticing
that \index{Bose-Fermi mixtures ! model} Bose-Fermi mixtures in optical
lattices have already been intensively studied
\cite{OspelkausS06,Gunter06,Best09}; these systems might
exhibit superconductivity due to \index{interactions ! boson mediated} boson-mediated fermion-fermion
interactions. Superexchange interactions \index{interactions ! superexchange}, demonstrated very recently in the context of
ultracold atoms in optical lattices \cite{Trotzky08}, are
believed also to play an important role in the context of
high-temperature superconductivity \cite{Lee06}. 
Remarkably, antiferromagnetic correlations 
for Fermion has been demonstrated in \cite{Greif13}
in dimerized lattices via radiofrequency band transfer and 
observed in momentum space via spin sensitive Bragg scattering of light \cite{Hart15}
and very recently also in situ via quantum gas microscopy \cite{Parsons16,Boll16,Cheuk16,Drewes16}.
Very recently antiferromagnetic 

\item{\it BCS-BEC crossover}. \index{BCS-BEC ! crossover} Physics of high $T_c$ superconductivity can be also
addressed with trapped ultracold gases. Weakly attracting
spin-$1/2$ fermions in such situations undergo at (very) low
temperatures the BCS \index{Bardeen-Cooper-Schrieffer (BCS) ! transition} Bardeen-Cooper-Schrieffer (BCS) transition to a
superfluid state of loosely bounded \index{Cooper pair} Cooper
pairs. Weakly repulsive fermions, on the other hand may form
bosonic molecules, which in turn may form at very low temperatures
a BEC \index{Bose-Einstein condensation (BEC) ! molecules}.
Strongly interacting fermions undergo also a transition to the
superfluid state, but at much higher $T$. Several groups  have
employed the technique of \index{Feshbach resonances} Feshbach
resonances \cite{Inouye98,Cornish00,Timmermans99} ---for a
recent review on this technique see \cite{Chin10}--- to
observe such BCS-BEC crossover. For the recent status of
experiments of the spin-balanced case see references in
\cite{Bloch08} and \cite{Giorgini08}---. There is
much more controversy regarding \index{spin ! imbalance mixtures}
imbalanced spin mixtures \cite{Zwierlein06,Partridge06a}.
First experimental signatures supporting pairing with finite
momentum in spin-imbalance mixtures, the so-called \index{Fulde-Ferrell-Larkin-Ovchinnikov (FFLO) !
state} FFLO state \cite{Fulde64,Larkin65}, have been observed
in one-dimensional Fermi gases \cite{Liao10}. Interestingly, topological nontrivial flat bands 
have been proposed for increasing the critical temperature of the superconducting transition \cite{Peotta15}.

\item{\it Frustrated antiferromagnets and spin liquids}.
The ``rule of thumb'' says that everywhere, in a vicinity of a
\index{high $T_c$ superconductivity} high $T_c$ superconducting
phase, there exists a (frustrated) antiferromagnetic phase.
Frustrated antiferromagnets have been thus in the center of
interest in condensed matter physics for decades. Particularly
challenging here is the possibility of creating novel, exotic
quantum phases, such as \index{quantum phases ! valence bond solids (VBS)} valence bond
solids, \index{resonating valence bond (RVB) states} resonating valence
bond states, and various kinds of quantum spin liquids ---spin
liquids of I and II kind \index{spin liquid ! I type} \index{spin
liquids ! II type}, according to C. Lhuillier
\cite{Misguich03,Lhuillier05}, and \index{spin liquid !
topological} topological and \index{spin liquid ! critical}
critical spin liquids, according to M. P. A. Fisher
\cite{Alet05a,Sachdev08}---. Cold atoms offer also in this
respect opportunities to create various frustrated spin models in
\index{lattice ! triangular} triangular, or even \index{lattice ! kagom\'e} kagom\'e lattices \cite{Santos04}. In the latter
case, it has been proposed by Damski \textit{et al}.
\cite{Damski05a,Damski05b} that cold \index{Fermi systems ! dipolar gas} dipolar Fermi gases\index{dipolar systems ! fermions}, or \index{Bose-Fermi mixtures ! model}
Bose-Fermi mixtures might allow to realize a novel state of
quantum matter: {\it quantum spin liquid crystal}, \index{quantum
spin liquid crystal} characterized by \index{N\'eel order} N\'eel
like order at low $T$ ---see also \cite{Honecker07}---,
accompanied by extravagantly high, liquid-like density of low
energy excited states.

\item{\it Topological order and quantum computation}. \index{topological ! order} \index{quantum computation ! topological} Several very ``exotic''
spin systems with topological order have been proposed recently
\cite{Kitaev06,Doucot05} as candidates for robust quantum
computing (for a recent review see \cite{Nayak08}). Despite their
unusual form, some of these models can be realized with cold atoms
\cite{Duan03,Micheli06}. Particularly interesting
\cite{Lewenstein06a} is the recent proposal by Micheli \textit{et
al}. \cite{Micheli06}, who propose to  use \index{molecules polar}
hetero-nuclear polar molecules in a lattice, excite them using
microwaves to the lowest rotational level, and employ strong
dipole-dipole interactions in the resulting spin model. The method
provides an universal ``toolbox'' for spin models with designable
range and spatial anisotropy of couplings. Experimental
achievements in cooling of heteronuclear molecules that may have a
large electric dipole moment \cite{OspelkausS06,OspelkausS09} have
opened possibilities in this direction. Recently, a gas of
ultracold ground state Potassium-Rubidium molecules has been
realized \cite{Ni08}.

\item{\it Systems with higher spins}. Lattice Hubbard models, or spin systems with higher spins are also related to
many open challenges; perhaps the most famous being the
\index{Haldane ! conjecture} Haldane conjecture concerning the
existence of a gap, or its lack for the
1D antiferromagnetic spin chains
with integer or half-integer spins, respectively. Ultracold spinor gases \cite{StamperKurn99,StamperKurn13} might help
to study these questions. Again, particularly interesting are in
this context spinor gases\index{spinor gases} in optical lattices
\cite{Imambekov03,Yip03a,Yip03b,Eckert07}, where in the
strongly interacting limit the Hamiltonian reduces to a
\index{Heisenberg ! generalized Hamiltonian} generalized Heisenberg
Hamiltonian. Spinor condensates in optical lattices have been addressed 
experimentally for ferromagnetic \cite{Widera05, Widera06, Becker10,Pedersen14} and antiferromagnetic \cite{Zhao15} interactions. 
Using Feshbach resonances \cite{Chin10} and
varying the lattice geometry \index{lattice ! geometry} one should be able in such systems to
generate a variety of regimes and quantum phases, including the
most interesting antiferromagnetic (AF) regime. Garc{\'i}a-Ripoll
\textit{et al}. \cite{Garcia-Ripoll04} proposed to use a
duality between the antiferromagnetic (AF) and ferromagnetic (F)
Hamiltonians, $H_{AF}=-H_F$, which implies that minimal energy
states of $H_{AF}$ are maximal energy states of $H_F$, and vice
versa. Since dissipation and decoherence are practically
negligible in such systems, and affect equally both ends of the
spectrum, one can study AF physics with $H_F$, preparing
adiabatically AF states of interest. Ytterbium fermionic atoms with N different 
spin states in an optical lattices implementing the fermionic SU(N) Hubbard model have 
been recently experimentally investigated \cite{Hofrichter15}.

\item{\it Fractional quantum Hall states}. \index{fractional quantum Hall effect (FQHE)} Since the famous work
of Laughlin \cite{Laughlin83}, there has been enormous
progress in our understanding of the fractional quantum Hall
effect (FQHE) \cite{Jacak03}. Nevertheless, many challenges
remain open like the direct observation of the anyonic character
of excitations or the observation of other kinds of strongly
correlated states. FQHE states might be studied with trapped
ultracold rotating gases \cite{Wilkin00,Cooper01}.
\index{artificial gauge fields ! rotating ultracold gas} Rotation induces
there effects equivalent to an ``artificial'' constant magnetic
field directed along the rotation axis. There are proposals about
how to detect directly fractional excitations in such systems
\cite{Paredes01}. Optical lattices might help in this task in
two aspects. First, FQHE states of small systems of atoms could be
observed in a lattice with rotating site potentials, or an array
of rotating microtraps ---cf. \cite{Popp04},
\cite{Barberan06},\cite{Dagnino07} and
\cite{Osterloh07} and references therein---. Second,
''artificial'' magnetic field might be directly created in lattices
via appropriate \index{artificial gauge fields ! tunneling control} control of the tunneling (hopping) matrix element in
the corresponding Hubbard model
\cite{Jaksch03}. Such systems will also allow to create FQHE
type states \cite{Mueller04,Sorensen05,Palmer06,Palmer08}.
Klein and Jaksch \cite{Klein09} have recently proposed to
immerse a lattice gas in a rotating Bose condensate; tunneling in
the lattice becomes then partially mediated by the phonon
excitations of the BEC and mimics the artificial magnetic field
effects. Last, but not least, a direct approach employing lattice
rotation has been developed both in theory
\cite{Bhat06,Bhat07}, and in experiments \cite{Tung06}.
The recent wave of very successful experiments creating artificial
or synthetic magnetic fields employing laser induced gauge fields
\cite{Y.-J.Lin09a,Y.-J.Lin09b} that are \index{artificial gauge
fields ! laser induced} achieved by using spatial
dependent optical coupling between different internal states of
atoms. Such approach is free from rotational restrictions.
Very recently, this successful wave have invested  also lattice experiments.
Paradigmatic 2D models displaying a topological insulating behavior like the Hofstadter \cite{Hofstadter76} and the Haldane \cite{Haldane88} models
have been realized experimentally in shallow harmonic traps in \cite{Aidelsburger13,Miyake13} and \cite{Jotzu14}, respectively,  
by exploiting superlattices and Bragg pulses. The Hofstadter model has been also realized in ladders and slabs, that to say in lattices with sharp
boundaries in one narrow dimension. Such dimension can be real as in the experiment by \cite{Atala14} or synthetic \cite{Celi14}, that to say 
with the rungs of the ladder formed by the internal spin states of atoms, which provide an extra dimension to the system \cite{Boada12}. 
The synthetic Hofstadter slabs has been realized experimentally for bosons \cite{Stuhl15} and fermions \cite{Mancini15} and the edge currents observed for the first time via spin-dependent 
measurements. Real and synthetic Hofstadter ladders and slabs offer a very promising route to the observation of many-body quantum Hall physics, both for bosons \cite{Petrescu13,Kolley15}
and fermions \cite{Barbarino15,Zeng15}.

\item{\it Lattice gauge fields}. Gauge theories, and in particular \index{lattice gauge theories (LGT)} lattice
gauge theories (LGT) \cite{Montvay97} are fundamental for
both high energy physics and condensed matter physics, and despite
the progress of our understanding of  LGT, many questions in this
area remain open (see e.g. ~\cite{Dalibard10}). Physics of
cold atoms might help here in two aspects: \index{artificial
gauge fields ! basic concepts} ``artificial''
non Abelian magnetic fields may be created in lattice gases via
appropriate control of the hopping matrix elements
\cite{Osterloh05} or in trapped gases using effects of
electromagnetically
induced transparency \cite{Ruseckas05}. One of the most
challenging tasks in this context concerns the  possibility of
realizing generalizations of \index{Laughlin ! state} Laughlin
states with possibly non Abelian fractional excitations. Another
challenge concerns the possibility of ``mimicking'' the dynamics of
gauge fields. In fact, dynamical realizations of \index{lattice gauge theories (LGT)} $U(1)$ Abelian gauge theory, that involve
\index{interactions ! ring exchange} ring exchange interaction in a
square lattice \cite{Buechler05}, or 3 particle interactions
in a \index{lattice ! triangular} triangular lattice
\cite{Pachos04,Tewari06} have been also proposed.
In the last four years the effort of simulating LGT in optical lattices
has received new impulse. Most of the attention has focused on gauge magnets or link models \cite{Horn81,Orland90,Chandrasekharan97},
which can be viewed as spin version of ordinary Hamiltonian LGT and, as such, are easier to be simulated. Both Abelian and 
non-Abelian LGT can be simulated by exploiting angular momentum conservation \cite{Zohar12,Zohar2013a}, SU($N$)-invariant interaction in earth-alkali like atoms \cite{Banerjee12,Banerjee13},
or the long-range interaction induced by Rydberg atoms \cite{Tagliacozzo13a,Tagliacozzo13b}. The proposed simulators would be  capable to probe the confinement of charges, for instance, by
measuring the string tension. Similar simulators have been proposed also for superconducting qubit \cite{marcos2013superconducting} and trapped ions \cite{hauke2013quantum}, 
where string breaking in Schwinger model has been very recently
experimentally demonstrated with four ions \cite{martinez2016real}.
These developments in quantum simulation has triggered also parallel developments in classical simulation both in 1D \cite{Banuls13,Rico14,Buyens14} and 2D or more \cite{Tagliacozzo14,Haegeman15}.

\item{\it Superchemistry}.  This is a challenge of quantum chemistry, rather than condensed matter physics:
to perform a chemical reaction in a controlled way, by using
\index{photoassociation} photoassociation or Feshbach resonances\index{Feshbach resonances}
from a desired initial state to a desired final quantum state.
Jaksch \textit{et al}.\ \cite{Jaksch02} proposed to use a Mott
insulator (MI) with two identical atoms, to create via
photoassociation, first a MI of homonuclear molecules \index{Mott
insulator (MI) ! molecular}, and then a molecular SF via ``quantum
melting'' \index{quantum melting}. A similar idea was applied to
heteronuclear molecules \cite{Damski02} in order to achieve
molecular SF. Bloch's group have indeed observed photoassociation
of \(^{87}\)Rb molecules in a MI with two atoms per site
\cite{Rom04}, while Rempe's group has realized the first
molecular MI using Feshbach resonances \cite{Volz06}.
Formation of three-body \index{Efimov effects} Efimov trimer
states was observed in trapped Cs atoms by Grimm's group
\cite{Kraemer06}. This process could be even more efficient
in optical lattices \cite{Stoll05}. An overview of the
subject of cold chemistry can be found in \cite{Krems09},
and in particular about cold Feshbach molecules in
\cite{Ferlaino09}. Control and creation of deeply bound
molecules in the presence of an optical lattice has been reported
in \cite{Danzl09}.

\item{\it Ultracold dipolar gases}.\index{dipolar systems ! basic concepts} Some of the most fascinating experimental and
theoretical challenges of the modern atomic and molecular physics concern ultracold dipolar quantum gases
---for reviews, see \cite{Baranov02},\cite{Baranov08b} and \cite{Lahaye09}---. The recent
experimental realization of the dipolar Bose \index{Bose-Einstein
condensation (BEC) ! dipolar}\index{dipolar systems ! bosons} gas
of Chromium \cite{Griesmeyer05}, and the progress in trapping and
cooling of dipolar molecules \cite{Ni08} have opened the path
towards ultracold quantum gases with dominant dipole interactions.
More recently, also dipolar gases of Dysprosium \cite{Lu11b} and Erbium \cite{Aikawa12}
have been cooled up to reach quantum degeneracy.  Very recently, dipolar gases \cite{dePaz13} and polar molecules \cite{Yan13} 
have been used to simulate quantum magnetism.
Dipolar BECs and BCS states of trapped gases are expected to
exhibit very interesting dependence on the trap geometry. Dipolar
ultracold gases in optical lattices, described by \index{Hubbard
model ! extended} extended Hubbard models, should allow to realize
various quantum insulating ``solid'' phases, such as \index{quantum
phases ! checkerboard (CB)} phase checkerboard (CB), and
superfluid phases such as \index{quantum phases ! supersolid (SS)}
supersolid (SS) phase \cite{Goral02a,Menotti07,Trefzger08}.
Particularly interesting in this context are the {\it rotating
dipolar gases} (RDG) \index{dipolar systems ! BEC}. Bose-Einstein
condensates \index{Bose-Einstein condensation (BEC) ! dipolar} of
RDGs exhibit novel forms of vortex lattices: square, ``stripe
crystal'', and ``bubble crystal'' lattices \index{lattice ! vortex}
\cite{Cooper05}. \cite{Baranov05} have demonstrated that the
pseudo-hole gap survives the large $N$ limit for the Fermi RDGs,
making them perfect candidates to achieve the strongly correlated
regime, and to realize  Laughlin liquid at filling $\nu=1/3$, and
\index{Wigner ! crystal} quantum Wigner crystal at $\nu\le 1/7$
\cite{Baranov08a} with mesoscopic number of atoms $N\simeq
50-100$.

\item {\it Wigner crystals or self-assembled lattices}. \index{Wigner ! crystal} Wigner, or \index{Coulomb ! crystal} Coulomb
type of crystals are predicted to be formed due to \index{dipolar
systems ! interactions}\index{interactions ! dipolar} long range
repulsive dipolar atom-atom or molecule-molecule interactions
\cite{Buchler07,Astrakharchik07,Astrakharchik08b}, or ion-ion
Coulomb interactions in the absence of a lattice \cite{Porras06a}.

\end{itemize}

Many of the above mentioned  challenges are discussed  discussed in the subsequent chapters of our  book \cite{Lewen2012}.
The interested reader should, however, extend her/his knowledge by turning toward
the excellent books of Pitaevskii and Stringari \cite{Pitaevskii03} and
Pethick and Smith \cite{Pethick08} on Bose-Einstein
Condensation in the weakly interacting regime, or to the books of
Fetter and Walecka \cite{Fetter03} or  X.G. Wen
\cite{Wen04} for many body and quantum field theory, among
others.

\section{Conclusions}
EU as well many other countries have concentrated recently considerable efforts to support quantum information science and technology. 
EU in particular will launch a Quantum Flagship, which will seek for technological advances in four pillars of the QI science; 
Quantum Computing and Simulations, Quantum Communications, Quantum Sensing and Metrology, and Quantum information Theory and Software.

While universal quantum computers remains still very challenging, quantum simulators already exist in the labs. 
Many of those are analogue or digital simulators of ``interesting'' quantum phenomena. 
In the recent year, however, a lot of effort has been devoted to quantum annealers, 
like D- wave machines that are designed to solve classical NP-complete or at least ultra-complex problems \cite{ lechner2015quantum, PhysRevA.93.052325, Raventos15}.  
Quantum annealers are perhaps the first quantum computing devices that have the chance to enter real technology and change our everyday life. 
In fact, a decisive progress toward definitive success of quantum computing devices may be achieved by integrating the different platforms \cite{kurizki2015quantum} 
that are best suited for different tasks. The development of such integrated structure would incarnate fully the spirit of the Quantum Flagship, and has quantum optics in its heart.   
Ones of the most intriguing applications of quantum simulators are nowadays facing towards problems that are not restricted to condensed matter, and even to physics {\it in stricto sensu} \cite{cirac2012goals}. 
Among them, simulation of Lattice Gauge Theories is especially fascinating and may lead to breaking the 20th century distinction between low and high energy physics. 
Such simulators may serve also for testing holographic principle \cite{zaanen2015holographic}. 
In fact, quantum simulation of high energy phenomena and gravitation \cite{barcelo2005analogue,boada2011dirac,rodriguez2016synthetic} 
is not the only way in which quantum optics and quantum technologies may have a revolutionary impact over our understanding of Nature at all scales. 
Indeed, in addition to the celebrate success of the interferometer LIGO in detecting gravitational waves \cite{abbott2016observation} 
as predicted by Einstein about a century ago \cite{einstein1916,einstein1918}, the impressive progresses in the optical clock standards \cite{ludlow2015optical} 
are expected  to lead in the next decade to direct tests of fundamental laws of physics by proving tighter and tighter bounds on the stability of fundamental constants \cite{safronova2014viewpoint}.

Thus, exciting time and challenges are expecting the quantum optics community in the following years that require a common effort.\\

Quantum workers of the world, unite!\\

\begin{verse}
This is  not the final struggle, but...\\
  Let us group together, and tomorrow\\
  The Quantum Internationale \\
  Will be the human race.
\end{verse}

We acknowledge financial support from John Templeton Foundation, European Union (EU IP SIQS, EU Grant PROACT QUIC, EU STREP EQUAM),
 the European Research Council (ERC AdG OSYRIS), and the Spanish MINECO (National Plan projects FOQUS -- FIS2013-46768 and Q-TRIP -- FIS2014-57460-P) and the Generalitat de Catalunya AGAUR (2014 SGR 874 and SGR2014-1639).

\bibliographystyle{apsrev4-1}


%

\end{document}